\documentclass[10pt,twocolumn,preprintnumbers,superscriptaddress,nofootinbib,aps,prl]{revtex4}

\usepackage{graphicx}
\usepackage{amsmath}
\usepackage{srcltx}
\usepackage[utf8]{inputenc}
\usepackage[colorlinks]{hyperref}
\usepackage{times}
\usepackage{amssymb}

\newcommand{\lsim}{\lesssim}
\newcommand{\gsim}{\gtrsim}
\newcommand{\eq}[1]{Eq.~(\ref{#1})}

\newcommand{\ord}[1]{\mathcal{O}{(#1)}}
\newcommand{\beq}{\begin{equation}}
\newcommand{\eeq}{\end{equation}}
\newcommand{\bea}{\begin{eqnarray}}
\newcommand{\eea}{\end{eqnarray}}

\newcommand{\Lam}{\Lambda}
\newcommand{\linv}{L^{-1}}

\newcommand{\appropto}{\mathrel{\vcenter{
  \offinterlineskip\halign{\hfil$##$\cr
    \propto\cr\noalign{\kern2pt}\sim\cr\noalign{\kern-2pt}}}}}

\begin{document}

\pagestyle{plain}

\title{\boldmath Ultraviolet-Infrared Bounds and Minimum Coupling in Effective Field Theories}

\author{Hooman Davoudiasl}
\email{hooman@bnl.gov}


\affiliation{High Energy Theory Group, Physics Department \\ Brookhaven National Laboratory,
Upton, New York 11973, USA}


\begin{abstract}
	
We provide a simple new argument for a lower bound on the coupling of a $U(1)$ gauge interaction in an effective field theory (EFT), originally obtained from the Weak Gravity Conjecture.  Our argument employs basic principles of quantum mechanical energy-time uncertainty and Lorentz invariance, plus Bekenstein's entropy bound on the ultraviolet (UV) and infrared (IR) scales of an EFT.  We show that using an alternative UV-IR relation based on the Cohen-Kaplan-Nelson (CKN) bound results in a stronger lower bound on the $U(1)$ coupling, consistent with the more stringent nature of the CKN relation.  Applicability of our reasoning to other interactions is briefly discussed.  We also slightly extend the CKN bound, by accounting for the effective degrees of freedom, and examine some of its phenomenological implications.

\end{abstract}
\maketitle


{\it Introduction}: Modern quantum field theory (QFT) has proven itself to be a predictive and systematic framework for understanding numerous physical phenomena, at microscopic distances.  Similarly, for over a century, General Relativity (GR) has been the state-of-the-art in describing gravitational interactions, from everyday distances to the largest structures in the Universe.  However, QFT has not been successfully combined with GR to allow a proper and predictive understanding of quantum gravity, though string theory appears to be a promising framework in this direction.      
Perhaps, once that work is complete, we will understand some of the puzzles which QFT seems to leave unresolved.  Of these, the cosmological constant problem - why the theory expectation for its size fails so overwhelmingly - is a glaring one \cite{Weinberg:1988cp}.  One is then tempted to ask whether what we seem to understand about gravity could indicate how QFT should be modified to get us closer to a more complete description of physical laws.           

In the above spirit, Cohen, Kaplan, and Nelson (CKN) made an intriguing proposal \cite{Cohen:1998zx} that drew on insights regarding properties of black holes \cite{Bekenstein:1973ur,Bekenstein:1974ax,Hawking:1974sw,Hawking:1976de,tHooft:1993dmi,Susskind:1994vu}.  They argued that a well-behaved effective field theory (EFT) cannot give rise to black hole states whose horizon size is larger than the length $L$ that defines the infrared (IR) cutoff scale $\sim \linv$ of the theory.  Since an EFT with ultraviolet (UV) cutoff $\Lambda$ is expected to be valid up to energy densities  $\ord{\Lambda^4}$, demanding that a spacetime region of size $L$ not include any such black holes then suggests \cite{Cohen:1998zx}  
\beq
L\, \Lambda^2 \lsim M_{\rm P}\,,\quad\text{(CKN)}
\label{CKN}
\eeq
where $M_{\rm P} \sim 10^{19}$~GeV is the Planck mass.  Here and in what follows we only consider the main parametric dependencies.  

As pointed out in Ref.~\cite{Cohen:1998zx}, for $\linv$ set by the Hubble rate $H_0\sim 10^{-33}$~eV \cite{Zyla:2020zbs}, the UV cutoff of the corresponding EFT is given by 
\beq
\Lambda\sim 3\times 10^{-3}~\text{eV}\,. \quad\text{(Cosmic EFT)}
\label{Cosmic-EFT}
\eeq 
This scale turns out to be near the inferred scale $\sim 10^{-3}$~eV of a cosmological constant, the simplest explanation of the ``dark energy"  responsible for the accelerated expansion of the Universe \cite{Zyla:2020zbs}.  In this interpretation, the QFT expectation for the value of vacuum energy setting the cosmological constant does not seem fine-tuned, which can possibly address a long-standing conceptual problem.

The CKN bound is stronger than that obtained from Bekenstein's requirement 
\beq
L \,\Lambda^3 \lsim M_{\rm P}^2\,,\quad\text{(Bekenstein)}
\label{Bekenstein}
\eeq
demanding the maximal entropy of a region of size $L$ be smaller than its surface area in fundamental units $\sim M_{\rm P}^{-2}$, {\it i.e.}, the entropy of a black hole of the same size.

Neither the CKN nor the Bekenstein bound involve any information about the type or strength of the interactions in the assumed EFT.  The derivation of the CKN relation invokes thermal ensembles up to temperatures $T\sim \Lambda$, corresponding to energy densities $\rho\sim \Lam^4$ \cite{Cohen:1998zx}, without specifying the interactions requited to form a thermal population.  Nonetheless, one may ask what conditions may be implied on the EFT interactions with the assumed UV and IR scales.  This is a question that we will examine in what follows.

When considering an EFT, one can typically estimate the largest energies to which it can extend, namely the UV cutoff $\Lam$.  This may be, for example, due to appearance of strong dynamics at high scales where  one loses control of the EFT, or a change of the theory resulting from spontaneous symmetry breaking.  Also, one may simply truncate the spectrum and ``integrate out" states with masses $\gsim \Lam$.  However, it is often less clear if and how one could set an IR length scale $L$ for the theory, beyond which the EFT is not applicable.  

In practice, one generally assumes that $L$ can be arbitrarily large.  However, this assumption is not consistent with the CKN or Bekenstein bounds in Eqs.~(\ref{CKN}) and (\ref{Bekenstein}), respectively.  That is, as $L\to \infty$ these bounds both imply that $\Lam\to 0$, corresponding to a vanishing range of validity and the collapse of the assumed EFT.  These UV-IR relations then require $L$ to have an upper bound for any reasonable value of $\Lam$.  But, is there a non-trivial {\it lower bound} on $L$? We attempt to address this question for long range forces acting on matter, corresponding to renormalizeable physical interactions.  For concreteness and to make contact with the results of Ref.~\cite{ArkaniHamed:2006dz} based on {\it Weak Gravity Conjecture (WGC)}, we will focus on a $U(1)$ gauge interaction, with coupling $g \lsim 1$ (so that it stays perturbative), acting on fermions of mass $m_f$ and charge $Q=1$.  We will briefly comment on other possibilities, later in our discussion. 

{\it Interactions in an EFT}: To go further, we ask: How does one establish, within an EFT, that there is a $U(1)$ gauge interaction?  This seems like a simple question, but it will lead to non-trivial consequences, as shown below.  If the $U(1)$ gauge interaction cannot be physically observable, it will be indistinguishable from a global symmetry.  Generally, global symmetries are taken to be incompatible with quantum gravity, since microscopic black holes, for example, can destroy global charges.  Thus, determining that the $U(1)$ corresponds to a gauge symmetry has physical implications as to whether the EFT could in principle be compatible with quantum gravity.            
                    
To make sure that the $U(1)$ is distinguishable from a global symmetry, it should correspond to dynamics.  With a UV cutoff at $\Lam$, the largest allowed energy associated with this gauge interaction is given by 
\beq
E_\Lam \sim g^2 \Lam\,.
\label{ELam}
\eeq 
One can think of the above as the potential energy $\ord{g^2/r}$ between two test charges separated by a distance $r\gsim 1/\Lam$.  Here, it is assumed that $r$ is small enough to be of order the UV cutoff length scale, but somewhat larger so that we do not exceed the validity regime of the EFT. 

According to the quantum uncertainty relation $\Delta E \Delta t \gsim 1$, one needs a time scale of order $\Delta t$ to measure energies of order $\Delta E$.  Hence, the smallest time scale corresponding to $E_\Lam$ is given by      
\beq
t_\Lam \gsim (g^2 \Lam)^{-1}.
\label{tLam}
\eeq

There are other ways to arrive at the above result.  In particular, an EFT must be able  to describe a thermal ensemble over the prescribed UV-IR range of its validity (as invoked in the original CKN derivation \cite{Cohen:1998zx}).  Let us then consider a plasma generated by the assumed $U(1)$ interactions with fermions, at temperature $T$.  The hard scattering time scale is given by $\sim (g^4 T)^{-1}$, corresponding to momentum transfers $q\sim T$.  However, one can show that the time scale for soft scattering is given by $t^s_T \sim (g^2 T)^{-1}$ \cite{Arnold:2002zm,Arnold:2003zc}, which corresponds to $q\sim g T$.  Since the thermal screening mass is $\ord{g T}$, any softer momentum exchange {\it does not} lead to a smaller time \cite{Arnold:2002zm} and hence $t^s_T$ is the shortest dynamical time scale.  In particular, for $T\sim \Lam$, we get the same estimate as in \eq{tLam}.  Hence, the EFT should {\it at least} allow for a time scale $t^s_\Lam\sim t_\Lam$, in order for the $U(1)$ interaction to have a physical effect through scattering. 

The above derivation assumed a relativistic plasma.  We now show that a non-relativistic medium does not lead to time scales shorter than $t_\Lam$.  The scattering cross section is given by $\sigma_f \sim g^4/q^2$ and hence the rate for scattering is set by 
\beq
\Gamma \sim (g^4/q^2)\, n_f
\label{Gamma}
\eeq 
where $n_f$ is the number density of charges, here assumed to be fermions of mass $m_f$.  The plasma frequency
\beq
\omega_p \sim g \sqrt{n_f/m_f}   
\label{omegap}
\eeq
sets the screening length and hence one cannot assume arbitrary low $q^2$, as in the relativistic case above.  Therefore, the maximum rate is given by
\beq
\Gamma_{max} \sim g^2 m_f.
\label{Gammax}
\eeq
Hence, if there is a time scale for scattering shorter than $t_\Lam$, one needs to have 
$g^2 m_f > t_\Lam^{-1}$.  This requires $m_f > \Lam$ which is not allowed in the EFT, since by definition $m_f < \Lam$.  

Another possible process in the non-relativistic limit is gauge boson scattering from fermions.  Here, the cross section is given 
\beq
\sigma_\gamma \sim g^4/m_f^2.  
\label{Gamg}
\eeq
The rate is again $\sim n_f \sigma_\gamma$.  Note that since we are assuming a non-relativistic setup, 
\beq
n_f \lsim m_f^3\,, 
\label{nfmax}
\eeq
since a higher density will lead to relativistic fermions: as the distance among fermions is $\ord{n_f^{-1/3}}$, their momenta would exceed their mass if $n_f$ exceeded the above limit.  In this case, we see that the maximum rate is given by $g^4 m_f$ which, as before, cannot be larger than $t_\Lam^{-1}\sim g^2 \Lam$.  Hence, we find that in either relativistic or non-relativistic case, $t_\Lam$ defines a minimal time scale to realize scattering and distinguish $U(1)$ from a global symmetry in a physical sense.

By Lorentz invariance, we identify $t_\Lam$ with a distance scale: $t_\Lam \sim L_\Lam$.  Based on the preceding arguments, the IR scale must then satisfy $L \gsim L_\Lam$ and, together with \eq{tLam}, we obtain        
\beq
g^2 \Lam \, L \gsim 1\,.\quad\quad \text{(ECR)}
\label{ECR}
\eeq 
For ease of reference, we will refer to the above as the Effective Coupling Relation (ECR).  One could find further motivation for \eq{ECR}, as will be presented in the Appendix. 

A possible question related to the identification of $t_\Lam$ with $L_\Lam$ in the above is whether one could have introduced a ``box" of size $L$ much smaller than $t_\Lam$ and wait for $t\gsim t_\Lam$ to detect the physical effects of our interaction.  However, this raises the question of how such a ``box" could confine the $U(1)$ (long range) gauge field, unless it is composed of states that are much more strongly coupled to that field than those assumed in the EFT.  This could introduce an inconsistency into the underlying assumptions in our discussion.  Hence, both by Lorentz invariance -- assumed here as a basic underlying principle -- and for consistency of definitions, one is motivated to take $t_\Lam$ as a good measure of the IR scale of the EFT.     

Before going further, let us comment on EFTs with multiple interactions $i=1,2,3,\ldots$.    Here, each interaction may have its own UV and IR cutoff, $\{\Lam_i,L_i\}$.  As discussed before, one often assumes $L_i\to \infty$, but we would like to consider the implications of UV-IR relations, Eqs.~(\ref{CKN}) and (\ref{Bekenstein}), that require finite $L_i$, $\forall i$.  To include several interactions in the same EFT, one must then take $\Lam=\min \{\Lam_i\}$.  Once the UV cutoff is set in this way,  \eq{ECR} requires $\linv \geq \max \{\linv_i\}$, such that the inequality is satisfied for any coupling.  

For simplicity, we will focus on one interaction, {\it i.e.} our assumed $U(1)$, since we would like to provide an independent argument for a bound on $g$, originally obtained via WGC \cite{ArkaniHamed:2006dz}\footnote{For a generalization of WGC to $N$ interactions see Ref.~\cite{Cheung:2014vva}.}.  In any event, it is necessary to know the proper UV and IR cutoffs for the $U(1)$, in order to understand the regime of validity of an EFT that contains it.  In what follows, we will further invoke a UV-IR relation to get a bound on $g$ in terms of $\Lam$.  We will show that \eq{ECR} can lead to a well-known result that was arrived at through very different considerations, in the context of WGC \cite{ArkaniHamed:2006dz}, as roughly summarized below.

{\it Coupling strength and UV cutoff}: In its simplest expression, the WGC posits that a necessary condition for EFTs to descend from a proper UV quantum theory of gravity, like string theory, is to exclude interactions that are weaker than gravity.  This was originally conjectured for the particular case of an Abelian $U(1)$ gauge theory with coupling $g$ \cite{ArkaniHamed:2006dz}.  One of the original arguments for the WGC is based on avoiding a large number of stable charged black hole remnants in a physical theory \cite{ArkaniHamed:2006dz}.  In Ref.~\cite{ArkaniHamed:2006dz}, it was argued that avoiding such an outcome requires a light particle of mass $m$ satisfying $m\lsim g M_P$ to be present in the physical spectrum.    
This inequality implies that the gravitational coupling strength $m/M_P$ must be smaller than $g$, making gravity the weakest force in the EFT.  Generalization of the preceding ``electric" WGC relation to magnetic monopoles was then shown to imply the UV scale for the theory must satisfy \cite{ArkaniHamed:2006dz} 
\beq
g \gsim \frac{\Lam}{M_{\rm P}}\,.\quad\quad\text{(WGC)} 
\label{WGC}
\eeq

We can obtain the above well-known conjecture for the UV cutoff scale of a $U(1)$ gauge interaction from the ECR.  To do this, one needs to eliminate $L$ from \eq{ECR}, which can be achieved via a UV-IR relation.   We use the Bekenstein condition in \eq{Bekenstein}, which together with \eq{ECR} immediately yields \eq{WGC}.  The use of Bekenstein condition is well-motivated in this context, since black hole entropy can be obtained in the context of string theory for special configurations (for some early work see, for example, \cite{Larsen:1995ss,Strominger:1996sh,Callan:1996dv}), and WGC is supported by string theory \cite{ArkaniHamed:2006dz} (for possible connections between the WGC and black hole entropy see, {\it e.g.}, Ref.~\cite{Cheung:2018cwt}).  Note, in particular, that \eq{WGC} obstructs obtaining a global symmetry for $g\to 0$, since it would require $\Lam \to 0$ \cite{ArkaniHamed:2006dz}.

Our argument for the WGC relation above involved three components: quantum mechanics, Lorentz invariance, and information about gravity.  The quantum aspect is very basic and, as stated before, is a direct result of energy-time uncertainty.  Lorentz invariance led us to associate time scales with distance scales.  However, the use of the UV-IR condition was motivated by agreement with the physics of black holes, which are well-established states in general relativity and hence in the low energy limit of string theory.  Thus, one could expect to find a result that is consistent with an EFT descending from string theory.  In fact, WGC was motivated by consistency with string theory, and we ended up with the same result in \eq{WGC}, but through an alternative route.    

So, what happens if one removes black holes from the EFT?  This is what the CKN bound was designed to achieve: to excise black holes from an EFT characterized by UV and IR cutoff scales $\Lam$ and $\mu = 1/L$, respectively \cite{Cohen:1998zx}.  Let us then consider using the CKN UV-IR relation of \eq{CKN} in conjunction with \eq{ECR}; we get    
\beq
g^2 \gsim \frac{\Lam}{M_{\rm P}}\,.\quad\quad \text{(MCC)}
\label{MCC}
\eeq
We will refer to the above \eq{MCC} as the Minimum Coupling Condition (MCC) for an EFT, to distinguish it from the condition (\ref{WGC}).  

Here, we would like to make a few remarks about the preceding results.  The condition in \eq{MCC} is stronger than that of \eq{WGC}.  This is as expected, since the CKN bound (\ref{CKN}) is stronger than the Bekenstein requirement (\ref{Bekenstein}).  As mentioned before, the WGC relation in \eq{WGC} is consistent with expectations from string theory.  For example, $g$ can be identified as the coupling constant of a $U(1)$ symmetry from closed heterotic strings compactified to a 4-dimensional EFT, with the string scale $M_{st}$ identified with the UV cutoff $\Lambda$ \cite{ArkaniHamed:2006dz} (or, in general, we can identify $g$ as the string coupling constant).  In that context, and with $\Lam \to M_{st}$,  the WGC bound (\ref{WGC}) is then saturated: 
\beq
g \, M_P \sim M_{st}.  
\label{strings}
\eeq
On the other hand, the MCC in \eq{MCC} -- derived here by using the CKN relation -- does not lead to the above and hence seems inconsistent with a string theory embedding of the EFT.  It seems plausible to trace this to the CKN requirement of excising black holes -- which are legitimate states in the low energy limit of string theory -- from the EFT.  Note that the Bekenstein UV-IR relation does not remove black holes from the theory and could in principle be consistent with a string EFT.  The above appears to imply that an EFT which obeys the CKN bound is incompatible with a string theoretical completion, though this conclusion may require further investigation beyond the scope of this work. 

{\it Other interactions}: We motivated the ECR in \eq{ECR} via very general principles, applied to EFTs.  Note that since the ECR was obtained by appeal to physics near the UV cutoff $\Lam$, it does not seem to require a massless mediator, as in an unbroken gauge symmetry, as long as the mediator mass $\ll \Lam$.  It thus appears that the ECR should apply equally well to a force mediated by a light scalar or vector boson.  Hence, our treatment seems to provide a generalization of the bounds (\ref{WGC}) and (\ref{MCC}) beyond $U(1)$ gauge interactions to other forces.      

So far, we have focused on non-gravitational EFTs.  However, if one can possibly extend the application of ECR to gravity as an EFT, one may identify $g\to E/M_{\rm P}$, where $E$ denotes energy, and $\Lam\to M_{\rm P}$, as appropriate for gravity.  Then, \eq{ECR} gives $(E^2/M_{\rm P}) L\gsim 1$.  Interestingly, if one takes $E$ to be the typical energy scale associated with the cosmic energy density $\rho\sim E^4$, one finds 
\beq
H^{-1} \lsim L\,,
\label{H}
\eeq
where $H\sim \sqrt{\rho}/M_{\rm P}$ is the Hubble scale, whose inverse sets the size of a causal patch of the Universe.  We can identify this as the ultimate IR scale or the ``box size" $L$ of the  theory.  Thus, in the case of gravitational interactions, a natural physical interpretation appears to require that the ECR be saturated.

{\it Phenomenology}: It is interesting to note that the CKN bound may be testable in precision measurements of the electron anomalous magnetic moment $a_e \equiv (g_e-2)/2$ \cite{Cohen:1998zx,Cohen:2021zzr}, as we will summarize here.  In Ref.~\cite{Cohen:1998zx}, CKN concluded that the possible effect from \eq{CKN} on $a_e$ scales like $(m_e/M_P)^{2/3}$, too small to be tested in experiments.  However, later work suggested a less suppressed parametric dependence $\sim \sqrt{m_e/M_P}$ \cite{Davoudi:2014qua,Banks:2019arz}.  Reference \cite{Cohen:2021zzr} has recently revisited the possibility of probing the CKN UV-IR bound in $a_e$ measurements and concluded that the effect $\sim \alpha/(2\pi) \sqrt{m_e/M_P}\sim 10^{-14}$, which may be accessible to experiments in the foreseeable future.
 
Here, we will slightly extend the CKN derivation \cite{Cohen:1998zx} to include $N_*$ degrees of freedom in the EFT, which would yield an energy density $\rho \sim N_*\,\Lam^4$ at the UV cutoff.  Using this expression, we obtain  
\beq
L \, \Lambda^2 \lsim M_P/\sqrt{N_*}\,.
\label{CKN*}
\eeq
We will assume this modified relation in the following.      
One can interpret the above dependence on $N_*$ as the running of $M_P$ \cite{Adler:1980bx,Dvali:2007hz,Calmet:2008tn}.  This effectively lowers $M_P$ which can significantly strengthen the bound for $N_*\gg 1$; in the SM and its typical extensions $N_*\sim 100$.  (For some of the phenomenological implications of very large numbers of states in particle physics and astrophysics, see Ref.~\cite{Davoudiasl:2020uig}.) 
 
Let us briefly consider the effect of the modified relation in \eq{CKN*} for electron $g_e-2$.  This is easily done by rescaling the result for $a_e \equiv (g_e - 2)/2$ obtained in Ref.~\cite{Cohen:2021zzr} which yields 
 \beq
 \frac{\alpha}{2\pi} \left(m_e \sqrt{N_*}/M_P\right) ^{1/2}.  
 \label{ae*}
 \eeq
Here, the inclusion of $N_*^{1/4} \sim 3$ increases the predicted CKN uncertainty slightly and makes it closer to the sensitivity of the currently most precise measurement $\delta a_e^{\rm exp} = 2.8\times 10^{-13}$ \cite{Hanneke:2008tm}.   
At the present time, the best predictions of $a_e$ differ by $\sim 1.4 \times 10^{-12}$, which corresponds to more than 5$\sigma$, due to a discrepancy in the measured values 
of $\alpha$ \cite{Parker:2018vye,Morel:2020dww}. However, the $a_e$ predictions do not deviate from experiment by more than 2.5$\sigma$.  We thus take the gravitational uncertainty to be $\lsim 10^{-12}$.  This suggest that the number of unknown degrees of freedom, potentially contributing to the running of $M_{\rm P}$, below $\Lambda \sim \ord{100~\text{GeV}}$ is roughly bounded by 
\beq
N_*\lsim 10^9.\quad\quad\text{(Theory Error on $a_e$)}
\label{g*}
\eeq 

One can also in principle include the effect of $N_*$ in the derivation of the Bekenstein bound \eq{Bekenstein}\footnote{For a discussion of black hole entropy in the presence of multiple species, see for example Ref.~\cite{Bekenstein:1994bc} and references therein.}
\beq
L \Lam^3 \lsim M_P^2/N_*
\label{Bekenstein*}
\eeq
Again, as may be expected, the inclusion of $N_*$ can be interpreted as the running of $M_P$, which strengthens the bound.  Accounting for this effect in our argument for obtaining \eq{WGC} then yields
\beq
g\gsim \frac{\Lam}{M_P} \sqrt{N_*}\,.
\label{WGC*}
\eeq

Before concluding, we make a comment about our reasoning that led to \eq{WGC}.  As discussed before, this lower bound on $g$ was originally derived in Ref.~\cite{ArkaniHamed:2006dz} for a $U(1)$ gauge interaction via theoretical arguments including those related to black hole remnants, expectations about a quantum theory of gravity, and hypothetical magnetic monopoles.  Our argument used quantum energy-time uncertainty, Lorentz invariance, and Bekenstein's bound.  Why these two seemingly very different approaches yield the same relation between $g$ and the UV cutoff scale of an EFT is an interesting question.  It may be a coincidence, but it appears plausible that there could be a way to relate the two paths for reaching the same result; we leave this as an open question here.    

{\it Summary}: 
In this work, we proposed a general relation - the Effective Coupling Relation (ECR) - among the UV and IR scales and the coupling of a $U(1)$ gauge theory.  The ECR was motivated as a basic quantum criterion on an observable gauged $U(1)$ that can be distinguished from a global symmetry in the EFT.  We showed that Weak Gravity Conjecture (WGC) bound on $U(1)$ gauge interaction strength $g$ can be obtained from the ECR, in conjunction with the Bekenstein UV-IR relation, related to black hole entropy.  We then showed that using the CKN UV-IR relation a different bound on $g$ can result that we referred to as the Minimum Coupling Condition (MCC).  The MCC is stronger than the WGC bound and also appears inconsistent with an EFT in the low energy limit of string theory.  Both features could be expected, as the CKN relation {\it (i)} represents a stronger condition on UV and IR scales than the Bekenstein relation and {\it (ii)} removes black holes, legitimate low energy states in string theory, from the EFT.  Our treatment can in principle extend to other dynamical forces, for example those mediated by a scalar.  We also introduced the effect of EFT degrees of freedom $N_*$ in the CKN bound, which can make it much stronger for $N_*\gg 1$.  The implications of the modified CKN bound for the electron magnetic moment measurements were also briefly discussed.  

\begin{acknowledgments}
This work is supported by the US Department of Energy under Grant Contract DE-SC0012704.  The author thanks Cliff Cheung for comments. 
\end{acknowledgments}

\appendix*

\section{Appendix}

\subsection{Further discussion of ECR}

Here, we will briefly discuss another way of arriving at \eq{ECR}.  Let us define a ``Bohr Radius" by 
\beq
R_{\rm B}(m_f) \sim (g^2 m_f)^{-1}.  
\label{RB}
\eeq
This sets the smallest bound state size of the theory and if the EFT is to provide a description for such a state, then one must have $R_{\rm B}(m_f) \lsim L$.  Since $\Lam \gsim m_f$, we then have $(g^2 \Lam)^{-1} \lsim L$, which is again the ECR obtained above.

\bibliography{uvir-ref}

\end{document}